\documentclass[twocolumn, switch]{article} % Method A for two-column formatting

\usepackage{preprint}
\usepackage{amsmath, amsthm, amssymb, amsfonts}

\usepackage[numbers,square]{natbib}
\usepackage{enumitem}
\usepackage[utf8]{inputenc}	% allow utf-8 input
\usepackage[T1]{fontenc}	% use 8-bit T1 fonts
\usepackage{xcolor}		% colors for hyperlinks
\usepackage[colorlinks = true,
            linkcolor = purple,
            urlcolor  = blue,
            citecolor = blue,
            anchorcolor = black]{hyperref}	% Color links to references, figures, etc.
\usepackage{booktabs} 		% professional-quality tables
\usepackage{nicefrac}		% compact symbols for 1/2, etc.
\usepackage{microtype}		% microtypography
\usepackage{lineno}		% Line numbers
\usepackage{float}			% Allows for figures within multicol

\usepackage{lipsum}		%  Filler text

\usepackage{newfloat}
\DeclareFloatingEnvironment[name={Supplementary Figure}]{suppfigure}
\usepackage{sidecap}
\sidecaptionvpos{figure}{c}

\usepackage{multirow}

\usepackage{titlesec}
\titlespacing\section{0pt}{12pt plus 3pt minus 3pt}{1pt plus 1pt minus 1pt}
\titlespacing\subsection{0pt}{10pt plus 3pt minus 3pt}{1pt plus 1pt minus 1pt}
\titlespacing\subsubsection{0pt}{8pt plus 3pt minus 3pt}{1pt plus 1pt minus 1pt}

\usepackage{tikz,xcolor,hyperref}

\definecolor{lime}{HTML}{A6CE39}
\DeclareRobustCommand{\orcidicon}{
	\begin{tikzpicture}
	\draw[lime, fill=lime] (0,0)
	circle [radius=0.16]
	node[white] {{\fontfamily{qag}\selectfont \tiny ID}};
	\draw[white, fill=white] (-0.0625,0.095)
	circle [radius=0.007];
	\end{tikzpicture}
	\hspace{-2mm}
}
\foreach \x in {A, ..., Z}{\expandafter\xdef\csname orcid\x\endcsname{\noexpand\href{https://orcid.org/\csname orcidauthor\x\endcsname}
			{\noexpand\orcidicon}}
}

\usepackage{xspace}

\newcommand{\figref}[1]{Figure~\ref{#1}}
\newcommand{\tabref}[1]{Table~\ref{#1}}
\newcommand{\secref}[1]{Section~\ref{#1}}


\title{Gen-C: Populating Virtual Worlds with Generative Crowds}

\fancypagestyle{firstpage}{
  \fancyhf{}
  \fancyfoot[C]{\small *correspondence: \texttt{a.panayiotou@cyens.org.cy}}

}

\usepackage{authblk}

\author[1,2\thanks{\tt{a.panayiotou@cyens.org.cy}}]{Andreas Panayiotou}
\author[2]{Panayiotis Charalambous}
\author[3]{Ioannis Karamouzas}

\affil[1]{Department of Computer Science, University of Cyprus}
\affil[2]{CYENS - Centre of Excellence}
\affil[3]{Department of Computer Science and Engineering, University of California, Riverside}

\begin{document}

\twocolumn[ % Method A for two-column formatting
  \begin{@twocolumnfalse} % Method A for two-column formatting

\maketitle
\thispagestyle{firstpage}

\begin{abstract}
Over the past two decades, researchers have made significant steps in simulating agent-based human crowds, yet most efforts remain focused on low-level tasks such as collision avoidance, path following, and flocking. As a result, these approaches often struggle to capture the high-level behaviors that emerge from sustained agent-agent and agent-environment interactions over time. We introduce \emph{Generative Crowds (Gen-C)}, a generative framework that produces crowd scenarios capturing agent-agent and agent-environment interactions, shaping coherent high-level crowd plans. To avoid the labor-intensive process of collecting and annotating real crowd video data, we leverage Large Language Models (LLMs) to bootstrap synthetic datasets of crowd scenarios. To represent those scenarios, we propose a time-expanded graph structure encoding actions, interactions, and spatial context. Gen-C employs a dual Variational Graph Autoencoder (VGAE) architecture that jointly learns connectivity patterns and node features conditioned on textual and structural signals, overcoming the limitations of direct LLM generation to enable scalable, environment-aware multi-agent crowd simulations. We demonstrate the effectiveness of our framework on scenarios with diverse behaviors such as a \emph{University Campus} and a \emph{Train Station}, showing that it generates heterogeneous crowds, coherent interactions, and high-level decision-making patterns consistent with the provided context.
\end{abstract}

\vspace{0.5cm}

  \end{@twocolumnfalse} % Method A for two-column formatting
] % Method A for two-column formatting

% \begin{figure*}[t]
%     \centering
%     \includegraphics[width=\linewidth]{Figures/Teaser.png}
%     \caption{Framework Overview: We leverage a Large Language Model to generate high-level crowd scenarios, generalize them into synthetic crowd data encoded as graphs, and learn a graph-feature space. Using input textual conditions, we sample this space to generate novel crowd scenarios.}
%     \label{fig:teaser}
% \end{figure*}

%\begin{multicols}{2} % Method B for two-column formatting (doesn't play well with line numbers), comment out if using method A

%%%%%%%%%%%%%%%  Main text   %%%%%%%%%%%%%%%
% \linenumbers
\section{Introduction}
\label{sec:introduction}

Simulating human crowds has long been an active research field, producing numerous models and tools for character behavior generation~\cite{Pelechano2016,Lemonari2022Authoring}. Several commercial solutions integrated into game engines and rendering software enable large-scale crowd rendering~\cite{golaem,mas,unity,unreal} in virtual scenes. While these approaches have advanced visual plausibility, they primarily target motion synthesis rather than the decision processes and interactions that govern behaviors over time.

Most agent-based methods focus on low-level navigation tasks such as collision avoidance, path following, and group steering~\cite{surveytoll,musse2021history}. While effective for local interactions, they struggle to model higher-level behaviors that arise as individuals pursue goals, interact socially, or respond to environmental cues. Examples include stopping to chat, browsing shop windows, or queuing for a train, activities that require planning and coordination but are rarely represented in current systems, leading to repetitive and less adaptive dynamics. Recent advances such as Meta’s WorldGen~\cite{wang2025worldgen} and Google’s Genie~\cite{bruce2024genie} highlight the need for scalable virtual worlds with coordinated agents.

In this work, we introduce \emph{Generative Crowds (Gen-C)}, a framework for synthesizing high-level behaviors of virtual agents for diverse multi-agent crowd simulations. Rather than targeting low-level physical fidelity, this work emphasizes the generation of plausible, diverse, and goal-driven high-level agent behaviors for autonomously populating complex environments.
At the core of our approach lies a \emph{crowd scenario graph}: a time-expanded graph that encodes the spatial and temporal evolution of the crowd, capturing agent-agent and agent-environment interactions.
With the advent of data-driven methods, an obvious approach is to learn such graphs directly from real data. Several works have modeled crowds by leveraging trajectories extracted from limited real-world datasets~\cite{Lerner2009Data,groupjehee,Lerner2009Fitting,greilcrowds,Panayiotou2025CEDRL}. However, the cost and effort required to collect and annotate such data, together with its limited coverage of high-level behaviors, quickly become a bottleneck and restrict generalization.

To reduce this dependency, we use Large Language Models (LLMs)~\cite{brown2020language} to automatically generate a small seed set of representative crowd scenarios. Although LLMs could directly script crowd behaviors, this approach scales poorly, demands extensive prompt design, and lacks efficiency for structured generation. Instead, we use LLMs only to bootstrap initial scenarios, represented as crowd scenario graphs. A generative model then learns a distribution of crowd actions and interactions from these graphs to synthesize new ones. Since existing graph embedding algorithms target static graphs and fail to model dynamic crowd scenarios, we introduce an architecture with two synergistic Variational Graph Autoencoders (VGAEs)~\cite{vgae} that jointly capture agent-agent and agent-environment dynamics. The trained model can be conditioned on natural language, enabling virtual crowd generation directly from text.

\begin{figure*}[t]
    \centering
    \includegraphics[width=\textwidth]{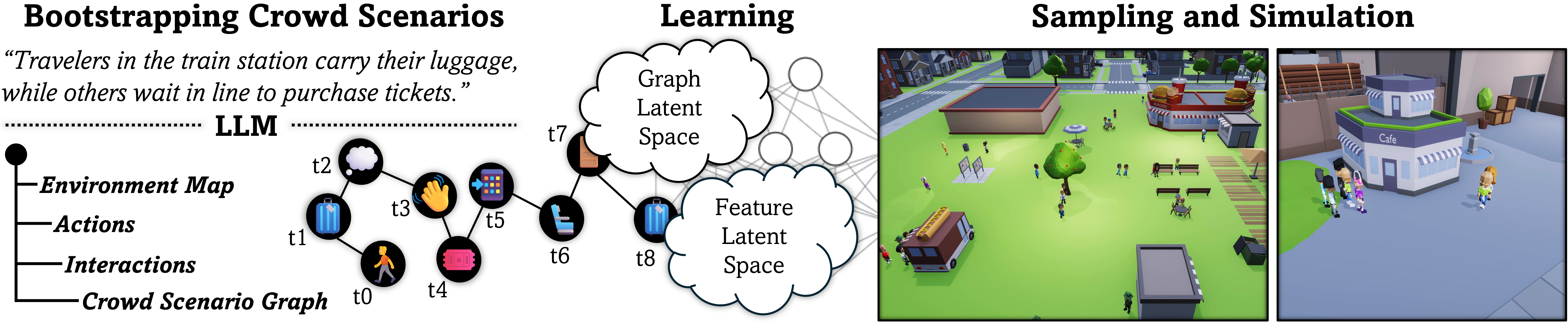}
    \caption{Framework Overview. We use a Large Language Model to generate synthetic datasets of crowd scenarios, which we represent as crowd scenario graphs to train a generative model. Conditioned on textual inputs, our model samples these graph representations to populate virtual worlds with diverse crowd scenarios.}
    \label{fig:teaser}
\end{figure*}

Our approach builds upon recent advances in generative AI, where models have been successfully applied to text-driven human motion synthesis~\cite{motionsurvey} and agent or robot planning~\cite{Liu2025Aligning}. However, these methods mainly focus on single-agent scenarios~\cite{wang2022humanise,tevet2023human,moconvq,motiongpt,wang2024sims}.
Closer to our work, text-based diffusion models have been explored for coordinating groups~\cite{ji2024text} and synthesizing crowd trajectories~\cite{rempe2023trace}, but they remain limited to unstructured motion patterns and a narrow range of behaviors. 
In contrast, Gen-C targets multi-agent settings and emphasizes the generation of diverse, high-level behavior compositions and interactions conditioned on scene context, enabling richer and more expressive characters in virtual scenes.
Overall, we introduce:
(i) a graph-based representation that captures temporal agent-agent and agent-environment interactions for virtual crowds; (ii) a text-conditioned dual-VGAE architecture that jointly learns graph structure and features, enabling scalable generation of novel multi-agent crowd scenarios; and (iii) a synthetic data pipeline, leveraging LLMs to bootstrap crowd scenarios, reducing reliance on real-world annotations.

\section{Related Work}
\label{sec:related-work}

\noindent\textbf{Crowd Simulation.}
The study of simulating crowds has a long history, with early works introducing rule-based models for navigation and flocking~\cite{Reynolds1999Steering}, or pedestrian activities~\cite{Shao2005Autonomous}, while more complex character behaviors, such as those in computer games, are often authored using state machines~\cite{yang2020review}.
Subsequent research proposed higher-level control mechanisms, such as crowd patches and sketch-based navigation fields~\cite{yersin2009crowdpatches,jordao2014sculpting,Patil2011Directing}, which allowed users to shape collective motion in structured environments. A major thrust of later work has been to increase plausibility by modeling \emph{heterogeneity}, for example through group-level properties~\cite{Ren2016Group}, individual personality traits~\cite{Durupinar;11,Kim2012Interactive,Guy2011Simulating}, or stylistic variations across different types of crowds~\cite{Ju2010Morphable}.
These methods highlighted the importance of diversity for believable simulations but typically required manual specification or extensive parameter tuning, motivating stochastic authoring approaches that provide direct user control over high-level crowd activity distributions~\cite{alin2014stochastic}.
In parallel, data-driven approaches emerged that learn behavioral patterns from real-world crowd trajectories~\cite{Lee2007Group,Lerner2007Crowds,Zhao2017CLUST,Lemonari2025MPACT}, sometimes adopting graph-based representations~\cite{Lai2005Group,Charalambous2014PAG} that connect crowd modeling to broader developments in machine learning. More recently, reinforcement learning and hybrid imitation-learning methods~\cite{Panayiotou2022CCP,Panayiotou2025CEDRL,rempe2023trace,greilcrowds}, diffusion-based generative approaches~\cite{huang2025diffusecloc,bae2025crowdes}, and graph network-based methods~\cite{Shi2023learning} have demonstrated the potential of scalable and adaptive policies that capture richer dynamics. Despite advances in crowd simulation, most existing methods focus on low-level navigation or local diversity, neglecting structured high-level multi-agent behaviors driven by agent-agent and agent-environment interactions. Our framework automatically generates heterogeneous, scenario-driven crowd behaviors capturing these multi-layered interactions.

%%%%%%%%%%%%%%%%%%%%%%%%%%%%%%%%%%%%%%%%%%%%%%%%%%%%%%%%%%%%%%%%%%%%%%%%
\noindent\textbf{Large Language Models and Autonomous Agents.}
LLMs have been shown to be effective in domains beyond Natural Language Processing (NLP) such as task planning, embodied reasoning, autonomous agents, and floor plan design~\cite{huang2022inner,song2023llmplanner,rana2023sayplan,Sun2023InteractivePU,Park2023GenerativeAgents,floorplan}.
Here, we explore an architecture that allows an LLM to generate a sequence of crowd behaviors, over time and space, that forms synthetic crowd datasets. We further exploit the power of generative NLP models to enable text-guided synthesis of high-level crowd behaviors. Along these lines a lot of work has explored generative models for text-driven human motion generation tasks~\cite{motionsurvey}. Despite impressive results, including generating scene-and-language conditioned motions~\cite{wang2022humanise,tevet2023human,moconvq,motiongpt,wang2024sims}, such methods typically focus on individual characters. But a human crowd is much more than that, consisting of multiple agents performing a sequence of actions while interacting with other agents and the environment. Along these lines, the recent work of~\citet{ji2024text} uses a text-based diffusion model combined with LLMs to steer groups of agents to predefined destinations, while~\citet{lemonari2024lexi} proposes a learning paradigm for mapping text to steering behavior parameters. Our work seeks to learn high-level behaviors captured via crowd scenario graphs that can complement the low-level steering capabilities of such agents.

%%%%%%%%%%%%%%%%%%%%%%%%%%%%%%%%%%%%%%%%%%%%%%%%%%%%%%%%%%%%%%%%%%%%%%%%
\noindent\textbf{Graph Generation.}
Beyond our domain, learning graph representations is fundamental to many real-world applications~\cite{gsurvey}. While much prior work has focused on static graphs using models like VGAEs~\cite{vgae}, recent efforts extend to dynamic graphs whose structures evolve over time. These include approaches using neural temporal point processes, attention mechanisms~\cite{dgsurvey}, and deep architectures such as Temporal Graph Networks~\cite{tgn_icml_grl2020} and Spatio-Temporal Graph Convolutional Networks~\cite{stgn} for modeling temporal dependencies and forecasting. While related, our work focuses on \emph{crowd scenario graphs}: dynamic graphs with rich relational features that capture the evolution of crowds. Unlike most dynamic graph models aimed at tasks such as link prediction, node classification, or recommendation, our formulation targets the evolution of crowd interactions. Closely related, conditional graph generation has also gained attention, with most approaches generating only the graph structure while assuming fixed node and edge features. For instance, NGG~\cite{Evdaimon2024neural} employs conditioned latent diffusion, GraphVAE~\cite{Simonovsky2018Graphvae} integrates conditional codes, and MOOD~\cite{Lee2023Mood} guides diffusion sampling with property gradients. Sequential models~\cite{Li2018Learning} use conditional inputs for nodes and edges initialization, while others embed conditions within message-passing schemes~\cite{Li2020Dirichlet}. Complementing these approaches, our framework jointly models graph structure and node features through separate, condition-aware latent distributions, enabling generation directly from text.

\section{Overview}
\label{sec:overview}

\noindent\textbf{Preliminaries.}
A \emph{crowd scenario graph} represents a \emph{crowd scenario}, consisting of agents, a series of events, environment locations, and the associated actions and interactions. To simplify the generation and learning process, we define a finite set of actions and locations which, when combined, can describe a wide range of crowd behaviors. Specifically, an \emph{Action} is a label from $\mathbf{\mathcal{A}ct}=~$\{wait, sit, wander, queue, enter/exit, wave at, discuss, meet, object interact, talk on phone, read, look at, carry\} that specifies an agent's behavior at a given time. When multiple agents perform a shared action (e.g., \emph{discuss}), this is referred to as an \emph{Interaction}. An \emph{Environment Location} is defined by its name, category, position, and scale. While the name may be abstract, 
the category is selected from a finite set:
$\mathbf{\mathcal{L}oc}=~$\{building, room, entrance, exhibit, furniture, outdoor area, item, service area\}. Finally, an \emph{Event} encodes a sequence of actions together with their location and participating agents. 
Specifically, each event consists of a set of \textit{action-location} pairs (e.g., \{discuss, room\}), the participating agents, and a trigger condition that determines its initiation (i.e., starts after a specific event).

%%%%%%%%%%%%%%%%%%%%%%%%%%%%%%%%%%%%%%%%%%%%%%%%%%%%%%%%%%%%%%%%%%%%%%%%
\noindent\textbf{The Gen-C Framework.}
Our framework for generating crowd scenarios consists of three main components (\figref{fig:teaser}): \emph{Synthetic Data Generation}, \emph{Learning}, and \emph{Scenario Generation}. First, we use an LLM to produce diverse synthetic scenarios through targeted queries that contains events describing agent actions, interactions, and environment placements (\secref{sec:quering-scenario}).
These scenarios are transformed into graph structures encoding action sequences, relationships, and interactions (\secref{sec:time-expanded-graph}).
Second, we train two synergistic VGAE models: 
one reconstructs the graph structure (agent interactions), while the other defines node features (agent actions and navigation). A text-conditioned prior integrates natural language, aligning the generated graphs with input descriptions (\secref{sec:learning}). Finally, novel scenarios are synthesized by sampling from the learned latent distributions, producing contextually consistent multi-agent crowd behaviors (\secref{sec:generating}).  

\section{Synthetic Data Generation}
\label{sec:data-generation}

Simulating diverse high-level crowd behaviors requires large amounts of high-quality data. However, collecting and annotating real-world crowd data is costly and labor-intensive, and existing crowd datasets are limited in scope, focus primarily on trajectory-level information, and lack high-level semantic annotations. To mitigate these limitations, we explore the use of recent advances in LLMs.

%%%%%%%%%%%%%%%%%%%%%%%%%%%%%%%%%%%%%%%%%%%%%%%%%%%%%%%%%%%%%%%%%%%%%%%%%
\subsection{Bootstrapping Crowd Scenarios via LLMs}
\label{sec:quering-scenario}

To generate simulations that capture representative crowd dynamics, we design two tailored LLM queries, each serving a specific purpose. The process begins with a single-sentence input $S_{in}$ describing a crowd scenario, which serves as the seed for generation. We construct a set of $100$ such seeds and, for each, create $50$ paraphrased variants to increase diversity, yielding $5000$ textual inputs in total per theme; in this work we introduce two themes, University Campus and Train Station. 
Each sentence encodes three main components: a location phrase, one or more subject noun phrases, and one or more action verb phrases, without following a fixed syntactic template. We avoid rigid structures to promote linguistic diversity and reduce the risk of overfitting to narrow prompt patterns. An example of $S_{in}$ is \emph{``Around the campus park, students sit to have lunch, talk with others, and wave at passing friends.''} Additional examples, along with exact prompts and parameters, are provided in the supplementary material. We use OpenAI’s \texttt{gpt-4.1}~\cite{openai2024gpt4o} and apply two sequential queries ($Q1$ and $Q2$), each constrained to return JSON-formatted outputs.

\emph{Generating the Environment ($Q1$).}
The first query prompts the LLM to generate a plausible environment layout from $S_{in}$, specifying \emph{Locations} relevant to the scenario. The output may include places like a ``coffee shop'', ``park'', or ``entrance'', together with their category, position, and scale.

\emph{Authoring Crowd Events ($Q2$).}
The second query defines the sequence of agent actions and interactions, both with the environment and with other agents. It takes as input $S_{in}$ together with a simplified list of the previously generated environment, containing only the location names for context. The output is a list of events, as previously defined in~\secref{sec:overview}.

By combining responses from both queries, we construct a high-level crowd scenario $CS$ from an $S_{in}$. We then iterate over the generated events to extract each agent’s actions and group memberships. As LLMs lack precise temporal detail, event sequences are derived from initiation constraints, and action durations are sampled from action-specific distributions. This yields a record $Rec_i$ for every agent $i$, capturing its actions and interactions throughout its lifespan in the scene.

\begin{figure*}[t]
    \centering
    \includegraphics[width=.85\linewidth]{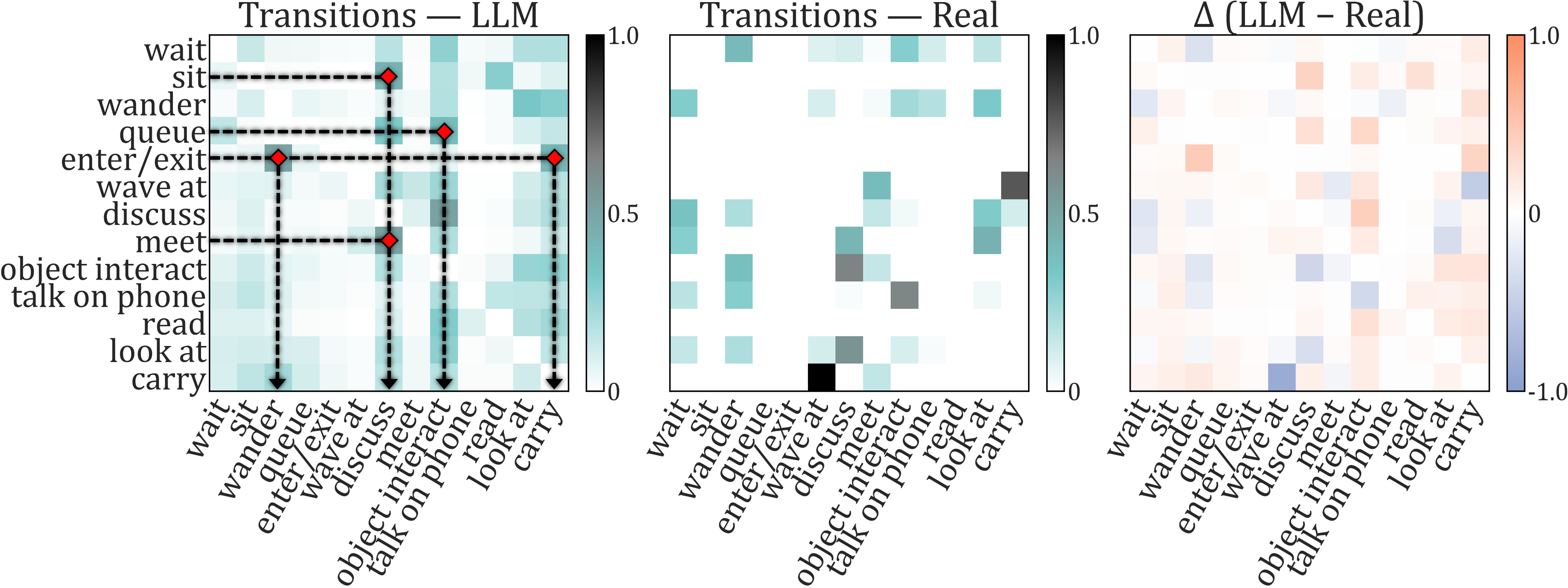}
    \caption{Per-agent action transitions from LLM generations and real-world annotations, and deviation between the two ($\Delta$) in Train Station theme. Rows represent the ``from'' action where columns the ``to''.}
    \label{fig:llm_vs_real}
\end{figure*}

\noindent\textbf{Assessment of LLM-bootstrapped Data.}
After using an LLM to generate crowd scenarios, we evaluate the reliability and fidelity of the results by annotating high-level action plans of individuals interacting in real train station environments, using a curated set of YouTube videos totaling approximately 30 minutes of footage. Comparable data for university campus environments are scarce, primarily due to ethical restrictions on video collection, and existing datasets such as UCY Dataset~\cite{Lerner2007Crowds} exhibit limited diversity in observed actions. Specifically, we divide each video clip into $5$-second windows and assign an action label from $\mathbf{\mathcal{A}ct}$ to each person visible in the frame.

In~\figref{fig:llm_vs_real} we report the action transition matrix obtained by the LLM bootstrapping on Train Station, real-world annotations, and the deviation between the two $\Delta$(LLM-Real). The LLM exhibits appropriate progressions for a Train Station—e.g., \textit{sit}→\textit{discuss}, \textit{queue}→\textit{object interact} (ticketing), \textit{enter}→\textit{wander}/\textit{carry} (luggage), and \textit{meet}→\textit{discuss}. The $\Delta$ panel shows small deviations indicating that temporal “what-comes-next” grammar is broadly aligned with annotated data, suggesting that LLM sequences follow coherent and logical patterns. We note that some actions are very rare or absent in the annotations (e.g., \textit{wave at}) due to the nature of the videos. Moreover, examples of elevated LLM rates include \textit{enter/exit}$\rightarrow$\textit{wander}, and \textit{read}$\rightarrow$\textit{talk on phone}, which likely arise from annotation constraints: the camera’s narrow field of view limits tracking of “wandering,” and \textit{talk on phone} is subtle and difficult to label reliably at distance or with restricted resolution/angles.

As an additional statistic, we measure diversity of action plans using the \emph{normalized entropy} of per-agent action sequences. Let $\mathcal{S}$ be the set of distinct action sequences and $p(s)$ their empirical frequencies. We compute $H(p) \;=\; -\sum_{s\in\mathcal{S}} p(s)\,\log p(s)$ and report the normalized value $\hat H = H(p)/\log|\mathcal{S}| \in [0,1]$: $\hat H=0$ means one pattern dominates; $\hat H=1$ means all patterns are equally likely. By this measure, LLM sequences closely match the annotated real-world data: $\hat H_{\text{LLM}}=0.949$ $(95\%\ \text{CI }[0.916,\,0.970])$ vs.\ $\hat H_{\text{Real}}=0.936$ $(95\%\ \text{CI }[0.935,\,0.939])$. The overlapping intervals indicate that LLM data follow real-world sequence variety for the given context.
Finally, it is worth noting that the reliability of synthetic LLM data should be interpreted at Gen-C’s level of abstraction: semantically plausible, context-aware high-level behavioral plans, rather than distributional equivalence to real-world trajectories. Accordingly, this evaluation should be viewed as a sanity check, although it is limited by its short duration and the partly subjective mapping to discrete action categories. More systematic real-world grounding through extensive evaluation is needed to fully establish the realism and naturalness of LLM-generated scenarios, which we identify as an important direction for future work.

%%%%%%%%%%%%%%%%%%%%%%%%%%%%%%%%%%%%%%%%%%%%%%%%%%%%%%%%%%%%%%%%%%%%%%%%
\begin{figure*}[t]
  \centering
  % --- Left Figure ---
  \begin{minipage}[b]{0.61\linewidth}
    \centering
    \includegraphics[width=\linewidth]{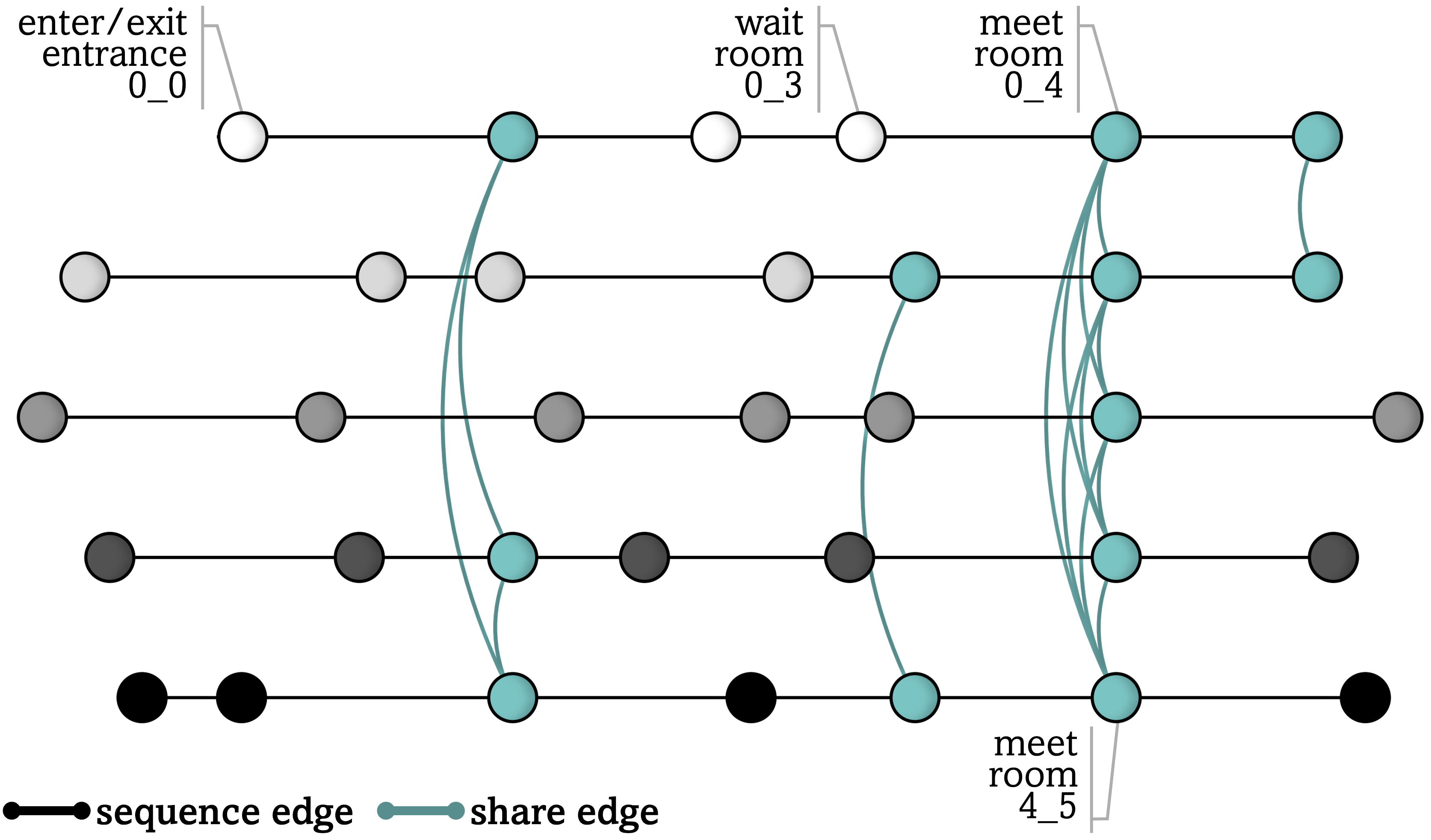}
    \caption{Snapshot of a crowd scenario graph. A node encodes an action, a location, and the agent ID along with a timestep as $i\_t$.}
    \label{fig:graph}
  \end{minipage}%
  \hfill 
  % --- Right Figure ---
  \begin{minipage}[b]{0.35\linewidth}
    \centering
    \includegraphics[width=\linewidth]{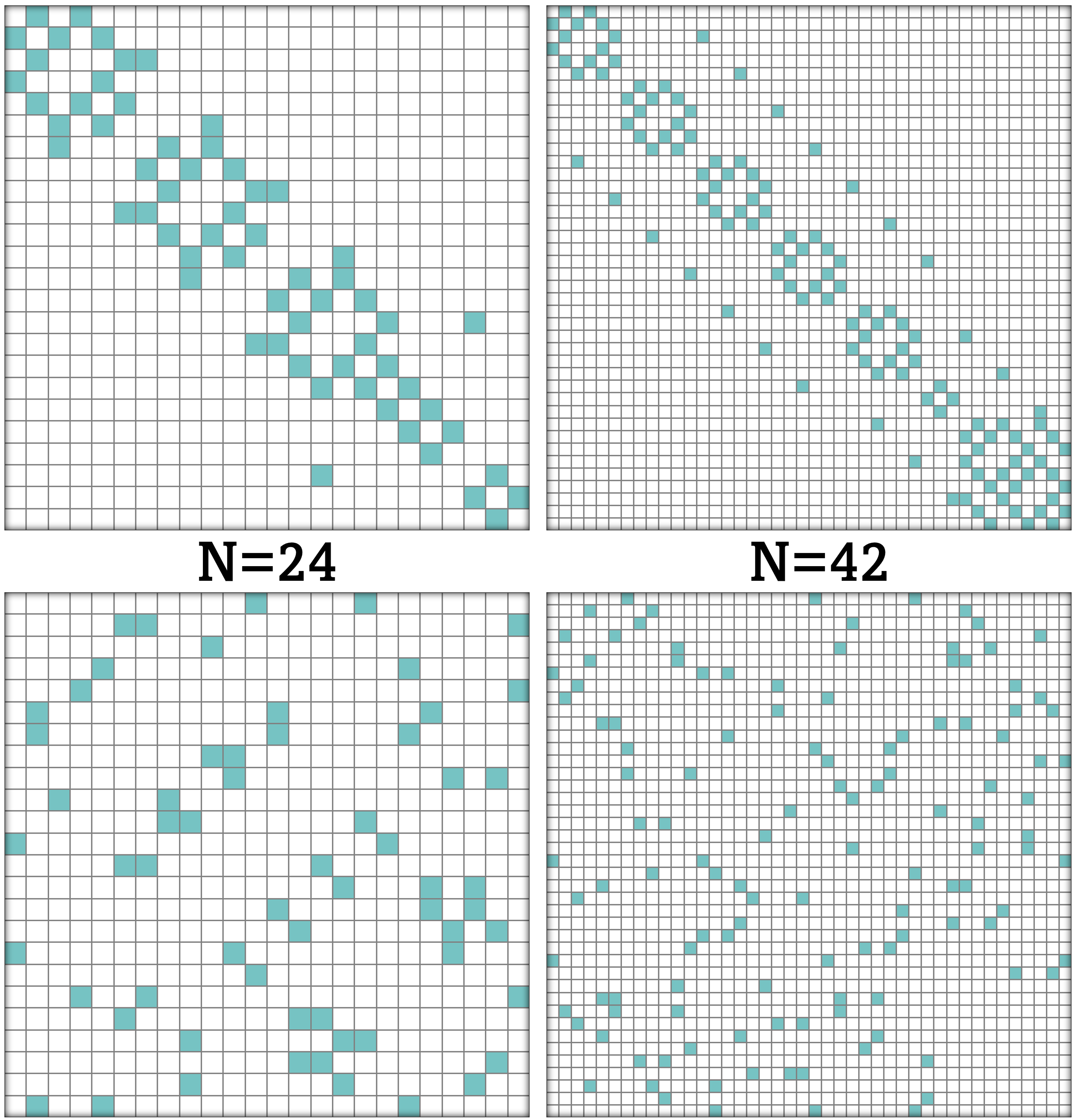}
    \caption{Canonical node ordering (top) vs. arbitrary node ordering (bottom).}
    \label{fig:node-ordering}
  \end{minipage}
\end{figure*}

\subsection{Graph-Based Representation}
\label{sec:time-expanded-graph}

Training a generative model for crowd scenarios requires an expressive data structure that captures agent-environment interactions and actions over time. We adopt a \emph{graph} representation that meets these criteria and introduce the crowd scenario graph, a time-expanded graph modeling dynamic crowd scenarios through temporally encoded actions and interactions.

As described in \secref{sec:quering-scenario}, during simulation we collect $Rec_{i}$ for each agent $i$. Iterating through $Rec_{i}$, we create a node $V_{i}^t = (i, A_{i}^t, L_{i}^t)$, where $t$ is the current timestep, $A_{i}^t \in \mathrm{A}ct$, and $L_{i}^t \in \mathrm{L}oc$; the combination of $A_{i}^t$ and $L_{i}^t$ defines an agent-environment interaction at $t$. We set the action sequence of $i$ using a ``sequence edge'' $E_{i}^t=(V_{i}^t, V_{i}^{t-1})$. Next, agent-agent interactions are captured by connecting nodes from different agents that share an action at a given $t$. A ``share edge'' $E_{i,j}^t=(V_{i}^t, V_{j}^t)$ is created for each pair of agents $i$ and $j$ interacting at $t$. At this stage, nodes of each agent $i$ are sequentially connected into a chain. If the chain has no ``shared edges,'' it forms an individual subgraph $H_k$ corresponding to agent $i$. When at least one ``shared edge'' exists, the chains of multiple agents merge into a subgraph representing a group of agents. Thus, for each $CS$, we construct an undirected graph $G=(V,E)$, consisting of multiple subgraphs $\mathcal{H} = \{H_1, H_2, \dots\}$, where $V$ is the set of nodes and $E$ the set of edges. Finally, each subgraph $H_k \in \mathcal{H}$ is treated as an individual training sample $(S_{in}, H_k, \text{agents}_k)$, where $H_k$ is the subgraph for group $k$, and $\text{agents}_k$ the agents in that group. \figref{fig:graph} presents a snapshot of a subgraph with five agents from a crowd scenario graph. 
\section{Learning Crowd Scenario Graphs}
\label{sec:learning}

While a synthetic LLM-generated crowd scenario graph, can drive a crowd of simulated agents, our goal is to generalize beyond manually querying an LLM whenever input conditions change or the simulation scale increases. To achieve this, we train a generative model on a dataset of crowd scenario graphs, enabling the synthesis of novel graphs directly from learned distributions.

%%%%%%%%%%%%%%%%%%%%%%%%%%%%%%%%%%%%%%%%%%%%%%%%%%%%%%%%%%%%%%%%%%%%%%%%
\subsection{Data Preprocessing}
\label{sec:training-preprocessing}

For each sample $(S_{in}, H_k, \text{agents}_k)$, we treat each subgraph $H_{k}$ as an individual undirected graph $G_k=(V_k, E_k)$, where $n=|V_k|$ is the number of nodes and $m=|E_k|$ the number of edges. The adjacency matrix $\mathcal{A}_{G_k} \in \mathbb{R}^{n \times n}$ is symmetric and encodes edge information: entries are $0$ if there is no edge between $v_i$ and $v_j$, $1$ if the edge is of type ``sequence,'' and $-1$ if it is a ``share'' edge.

\emph{Node Ordering.}
We enforce a canonical ordering of nodes to ensure consistency when constructing adjacency matrices. Different permutations of node indices can lead to inconsistency and higher training complexity, particularly for graph reconstruction tasks. To address this, we define a custom ordering where nodes are sorted first by event index, then by agent identifier (ID), and finally by the sequence index within the agent's actions, ensuring a consistent representation across samples. \figref{fig:node-ordering} shows sample adjacency matrices with canonical and arbitrary orderings. Note that this ordering standardizes node indexing for stable training but does not directly enforce logical action sequences, which are instead captured implicitly through the LLM-bootstrapped training data.

\emph{Node Features.}  
Each node is represented by a dense feature vector encoding action, location, agent ID, normalized position in the agent timeline, global sequence index, and Laplacian~\cite{Chung1997Spectral} positional encodings. Categorical features are mapped to learnable embeddings of size $16$, $8$, and $16$, respectively. The normalized time index contributes one scalar, the global index a 4D sinusoidal embedding, and the top-4 eigenvectors of the normalized Laplacian provide structural context, yielding a $49$-dimensional node representation.

\emph{Graph-Level Conditioning.}  
Beyond node-level features, we compute global signals that capture the context of a crowd scenario and serve as conditioning inputs for our model (see~\secref{sec:training}). Specifically, we include the normalized counts of nodes, agents, and events in each graph, using maximum values of $100$, $15$, and $30$ respectively. We also calculate an action-frequency vector that summarizes the distribution of action classes. These scalars and the action-frequency profile are concatenated with the embedding of $S_{in}$, obtained from the pretrained SentenceTransformer \texttt{all-mpnet-base-v2}~\cite{Reimers2019Sentence}, and jointly form the condition vector $C$.

\begin{figure*}[t]
    \centering
    \includegraphics[width=\linewidth]{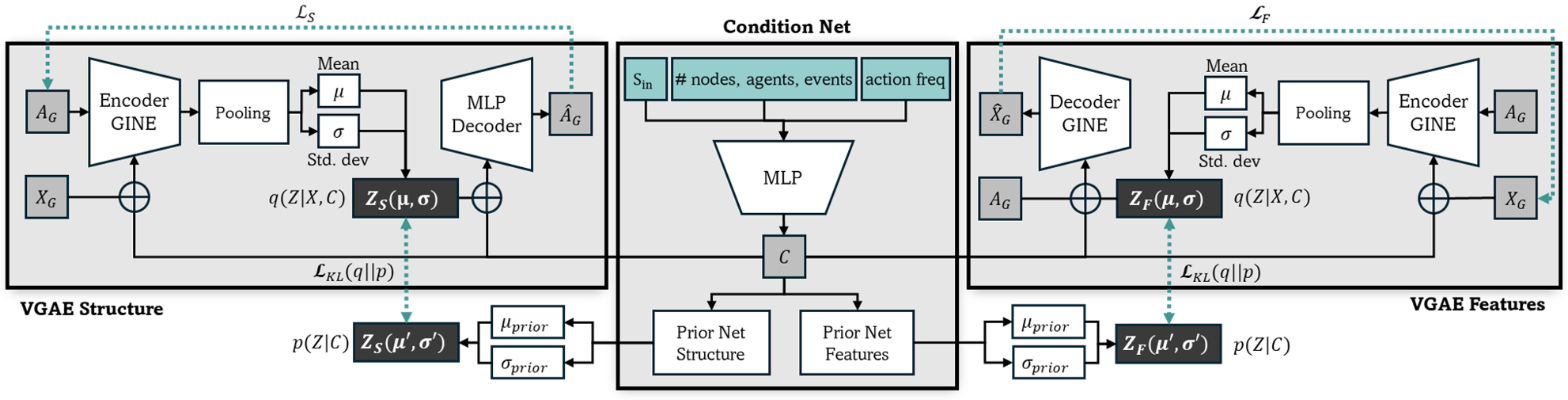}
    \caption{Our architecture comprises two synergistic VGAEs: one for graph reconstruction and one for node feature reconstruction, with encoders mapped to latent spaces regularized by a shared Condition Network.}
    \label{fig:training-architecture}
\end{figure*}

%%%%%%%%%%%%%%%%%%%%%%%%%%%%%%%%%%%%%%%%%%%%%%%%%%%%%%%%%%%%%%%%%%%%%%%%
\subsection{Training}
\label{sec:training}

We train two conditional VGAEs that jointly learn to reconstruct graph structure and node features from crowd scenario graphs; we present the model's architecture in~\figref{fig:training-architecture}.
The two models share a common encoder that produces latent embeddings, while using distinct decoders specialized for structure (VGAE-S) and features (VGAE-F). The encoder learns an approximate posterior distribution, $q(Z\mid X,C)$, where $Z_S$ and $Z_F$ are graph-level latents for structure and features, respectively. We observe that using a fixed standard normal prior, $p(Z) = \mathcal{N}(0, I)$, leads to a mismatch between $q(Z \mid X, C)$ and $p(Z)$, resulting in posterior collapse that degrades generalization. To mitigate this mismatch, we introduce a prior network that parameterizes the conditional priors $p(Z_S \mid C)$ and $p(Z_F \mid C)$, where $C$ is the condition vector. This conditional prior aligns the posterior with contextual and structural cues, enabling generation of graphs consistent with given conditions.

\noindent\textbf{Encoder.}  
Both VGAE-S and VGAE-F share the same encoder architecture, which maps an input graph $G=(V,E)$ with adjacency $\mathcal{A}_G$ and node features $\mathbf{X}_G$ into a latent representation. The encoder stacks multiple GINE layers~\cite{Hu2020Pretraining}, each implemented as a message-passing step followed by a residual connection. At layer $k$, node embeddings are updated as:
\begin{equation}
\mathbf{h}_v^{(k)} = \mathbf{h}_v^{(k-1)} + \text{GINE}^{(k)}\!\big(\mathbf{h}_v^{(k-1)}, \{(\mathbf{h}_u^{(k-1)}, e_{uv}) : u \in \mathcal{N}(v)\}\big),
\end{equation}
where $\mathbf{h}_v^{(k)}$ is the representation of node $v$, $\mathcal{N}(v)$ its neighbors, and $e_{uv}$ the edge feature (“sequence”/“share”). After message passing, node embeddings are aggregated with global add pooling into a single vector, normalized, and concatenated with the condition embedding $C$. Fully connected layers project this joint representation into the mean and log-variance of the Gaussian posterior,  
\begin{equation}
q(Z \mid X, C) = \mathcal{N}\Big( \mu(X, C), \operatorname{diag}\big( \exp(\log \sigma^2(X, C)) \big) \Big),
\end{equation}
where $\log\sigma^2(X,C)$ is the predicted log-variance and $\exp(\log\sigma^2)$ recovers a diagonal covariance with entries given by per-dimension variances. This parameterizes the latent distributions $q(Z_S \mid X,C)$ and $q(Z_F \mid X,C)$ for structure and features, respectively.

\noindent\textbf{Structure Decoder.}  
The structure decoder reconstructs the adjacency matrix $\widehat{\mathcal{A}}_G$ from the latent variable $Z_S$ and condition vector $C$. The concatenated vector $[Z_S;C]$ is processed by an MLP with nonlinear activations, and its outputs are mapped through a $\tanh$ function to produce edge values in $[-1,1]$. These values are assembled into a symmetric adjacency matrix,  
\begin{equation}
\widehat{\mathcal{A}}_G = U + U^\top, \quad U_{ij} = \tanh\!\big(f_\theta([Z_S;C])_{ij}\big)\;\;\text{for } i<j,
\end{equation}
with a zero diagonal. The sign of each entry encodes the edge type (“sequence” or “share”), while the magnitude reflects confidence.

\noindent\textbf{Feature Decoder.}  
The feature decoder reconstructs node features $\mathbf{X}_{G'}$ from the latent variable $Z_F$, the condition vector $C$, and graph-level positional encodings. At the node level, $Z_F$ and $C$ are expanded and concatenated with projected positional encodings, which combine Laplacian embeddings from the share-only subgraph (capturing the connectivity in individual groups induced by ``share'' edges) and the full graph (capturing the overall structure), together with a normalized group identifier. The fused representation is processed by stacked GINE layers with pre-normalization and residual connections, enabling message passing over the graph. To enforce consistency among agents that interact, node embeddings are collapsed into connected components defined by share edges, ensuring that members of the same group, at same timestep, are assigned a common action and location. Class logits are predicted at this group level and then uncollapsed back to the node level for supervision. Formally, for each node $v \in V$, the decoder outputs categorical distributions over actions and locations as: $\widehat{\mathbf{X}}^{\text{act}}_v = \text{softmax}(f^{\text{act}}_\theta(h_v))$, and $\widehat{\mathbf{X}}^{\text{loc}}_v = \text{softmax}(f^{\text{loc}}_\theta(h_v))$,
where $h_v$ is the reconstructed embedding of node $v$ after the collapse--predict--uncollapse pipeline. These distributions are directly compared with the ground-truth labels $\mathbf{X}_v^{\text{act}}$ and $\mathbf{X}_v^{\text{loc}}$ during training.

\noindent\textbf{Condition Network.}  
Both VGAE models use the same condition network, which encodes the text embedding of $S_{in}$ together with normalized counts of nodes, agents, and events, as well as an action-frequency vector, to produce the condition vector $C$. This vector is also used to train two prior networks that parameterize the conditional priors $p(Z \mid C)$ for VGAE-S and VGAE-F, yielding their respective means $\mu_{\text{prior}}$ and log-variances $\log\sigma_{\text{prior}}^2$. The condition vector is integrated into both the encoding and decoding phases. In the encoder, $C$ is concatenated with the pooled graph representation to produce the posterior distribution $q(Z\mid X,C)$, parameterized by $\mu$ and $\log\sigma^2$. During decoding, it is concatenated with the latent variable to guide reconstruction.

\noindent\textbf{Training Objective.}
VGAE models maximize the ELBO~\citep{Kingma2013AutoEncodingVB}, balancing reconstruction fidelity with latent space regularization. VGAE-S uses smooth L1-error (SmoothL1) loss $\mathcal{L}_S$ to compare predicted adjacency $\widehat{\mathcal{A}}$ against ground-truth $\mathcal{A}$, while VGAE-F applies node-level cross-entropy (CELoss) $\mathcal{L}_F$ for action and location predictions. The losses are defined as:
\begin{equation}
\begin{split}
\mathcal{L}_S &= \sum_{i,j} \text{SmoothL1}\!\left(\mathcal{A}_{ij}, \widehat{\mathcal{A}}_{ij}\right), \\
\mathcal{L}_F &= \sum_{v \in V} \Big( \gamma\text{CELoss}\big(\mathbf{X}_v^{\text{act}}, \widehat{\mathbf{X}}_v^{\text{act}}\big) + \text{CELoss}\big(\mathbf{X}_v^{\text{loc}}, \widehat{\mathbf{X}}_v^{\text{loc}}\big) \Big).
\end{split}
\end{equation}
where $\gamma=1.5$ is a balancing term.
In both models, the latent variables are regularized with a KL divergence term that penalizes deviation of the posterior from the conditional prior:  
\begin{equation}
\mathcal{L}_{KL} = \mathbb{E}_{z \sim q(Z \mid X,C)} \left[ \log q(Z \mid X,C) - \log p(Z \mid C) \right].
\end{equation}
The final training objectives are: $\mathcal{L}_{VGAE-S} = \mathcal{L}_S + \beta_S \,\mathcal{L}_{KL_S}$ and $
\mathcal{L}_{VGAE-F} = \mathcal{L}_F + \beta_F \,\mathcal{L}_{KL_F}$, where $\beta_S$ and $\beta_F$ are annealing factors that control the relative weight of the KL divergence.
\section{Crowd Scenario Generation}
\label{sec:generating}
At inference time, novel crowd scenarios are generated by sampling from the learned priors using the input condition $C$. The condition includes: a textual description $S_{in}$, graph-level statistics (counts of nodes, agents, and events), and a frequency map of the available actions. From this condition, the structure latent $Z_S \sim p(Z_S \mid C)$ and the feature latent $Z_F \sim p(Z_F \mid C)$ are drawn and decoded into an adjacency matrix $\widehat{\mathcal{A}}$ and node features $\widehat{\mathbf{X}}$, respectively. This ensures that both connectivity and node attributes are contextually aligned with the full conditioning signal.  
Each generated graph corresponds to a subgraph that encodes the actions and interactions of a group of agents, as introduced in~\secref{sec:time-expanded-graph}; the framework scales naturally to arbitrary numbers of agents or groups. From the decoded graphs, we extract per-agent sequences of actions, locations, and interactions. To instantiate these scenarios in a virtual scene, the environment's layout can either be provided manually or synthesized with a lightweight LLM query ($Q1$) (see~\secref{sec:quering-scenario}).
\section{Experiments and Evaluation}
\label{sec:evaluation}

We train on two synthetic crowd datasets, \emph{University Campus} and \emph{Train Station}. Following the procedure described in~\secref{sec:quering-scenario}, we generate 500 scenarios per dataset and convert them into graphs as in~\secref{sec:time-expanded-graph}, obtaining $22.9$K subgraphs for University Campus and $25.1$K for Train Station. Each dataset is split into 80\% training, 15\% validation, and 5\% testing. The source code, datasets, and trained models can be accessed at \url{https://github.com/apanay20/Gen-C}.

For training (see~\figref{fig:training-architecture}), the encoder uses two GINE layers with hidden size 128, each followed by dropout ($p=0.2$). Both VGAE-S and VGAE-F use a 32-dimensional latent space. The graph decoder is a three-layer MLP with hidden size 128, while the feature decoder uses two GINE layers (hidden size 128) followed by two linear heads predicting 13 action classes and 8 location classes. The condition network projects text embeddings (768D), scalar inputs (3D), and action frequencies (13D) to 128 dimensions each; their concatenation is processed by an MLP to produce a 32-dimensional condition vector, which the prior network maps to prior distribution parameters through two linear layers. LeakyReLU is used throughout, and LayerNorm is applied in the encoder, both decoders, and condition network for stability. Models are trained for 500 epochs with batch size 128, learning rate $5\times10^{-4}$ linearly decayed to $1.25\times10^{-4}$, and Adam with weight decay $3\times10^{-4}$. The $\beta_S$ and $\beta_F$ coefficients are cyclically annealed over 100-epoch cycles in the ranges $[0.001,3]$ and $[0.001,1]$, respectively. Training each model takes about one hour on a single NVIDIA RTX 4070 Ti GPU.

%%%%%%%%%%%%%%%%%%%%%%%%%%%%%%%%%%%%%%%%%%%%%%%%%%%%%%%%%%%%%%%%%%%%%%%%
\subsection{Ablation Study}
\label{sec:ablation}

To evaluate the impact of our design choices, we compare the proposed Gen-C model against two variations of our architecture on the test set: (i) a version trained without canonical node ordering (w/o Canonical), and (ii) a single VGAE model jointly trained on both structure and features (Single). Additionally, we include a constrained random baseline (Random), designed to support fair comparisons by generating graphs that conform to the same structural constraints as the ground truth (e.g., each node has 1-2 temporal edges). Node actions and locations are sampled from empirical priors and perturbed with mild noise, preserving marginal distributions while introducing controlled variability. Performance is reported using the KL divergence (KLD) between the distributions of ground-truth and generated samples across both structural and semantic measures. Lower values indicate closer alignment between generated and real samples. Structural fidelity is assessed using standard graph descriptors~\cite{Milena2021Graph}: the \emph{degree} of each node $v$ is its number of incident edges; the \emph{clustering coefficient} quantifies local transitivity, defined as $c(v) = \tfrac{2T(v)}{k_v(k_v-1)}$ where $T(v)$ is the number of triangles involving $v$ and $k_v$ its degree (with $c(v)=0$ if $k_v<2$); the \emph{graph diameter} is the maximum shortest-path length between connected nodes; and the \emph{average path length} is the mean shortest-path distance across all connected node pairs. Semantic alignment is evaluated by comparing the distributions of predicted and ground-truth action and location labels.

As shown in~\tabref{tab:ablation}, Gen-C consistently achieves the lowest divergence across most metrics in both datasets. The variant without canonical ordering performs poorly on structural measures, highlighting the importance of consistent node indexing for stable learning. The single VGAE model improves over this baseline but underperforms Gen-C, especially on semantic metrics, suggesting that separating structure and feature learning yields more accurate alignment between agent interactions and their actions/locations; interestingly, the diameter statistic benefits from joint training. The random baseline, while constrained, performs poorly on graph-structure metrics but shows respectable feature-level alignment, since actions and locations are sampled from the input distributions. Overall, the ablation confirms that the full framework is necessary to faithfully capture both structural and behavioral properties of crowd scenarios.

\begin{table*}[t]
  \caption{KL divergence (KLD) to ground-truth distributions.}
  \label{tab:ablation}
  \footnotesize%
  \centering%
  \setlength{\tabcolsep}{5.15pt}
  \begin{tabular}{l|cccc|cccc}
    \toprule
    \multirow{2}{*}[-0.1ex]{\bf Metric} & \multicolumn{4}{c|}{\bf University Campus (KLD) $\downarrow$} & \multicolumn{4}{c}{\bf Train Station (KLD) $\downarrow$} \\
    & w/o Canonical        & Single & Random & Gen-C & w/o Canonical & Single & Random & Gen-C\\
    \midrule
    Degree                 & 3.4955     & \it 0.0616 & 0.5953 & \bf 0.0560 & 3.2891     & \it 0.0510 & 0.4680 & \bf 0.0332 \\
    Clustering Coefficient & 0.9550     & \it 0.0341 & 0.2212 & \bf 0.0308 & 0.1472     & \it 0.0529 & 0.4381 & \bf 0.0384 \\
    Diameter               & 3.4630     & \bf 0.1448 & 0.4942 & \it 0.1559 & 3.6493     & \bf 0.0937 & 0.8044 & \it 0.1038 \\
    Average Path Length    & 0.9234     & \it 0.2761 & 0.8854 & \bf 0.1466 & 0.7142     & \it 0.1436 & 1.2349 & \bf 0.0990 \\
    Action                 & \it 0.1800 & 0.3579     & 0.2251 & \bf 0.1773 & \it 0.1280 & 0.3399     & 0.3681 & \bf 0.1057 \\
    Location               & \it 0.0242 & 0.0576     & 0.0244 & \bf0.0171  & \it 0.0188 & 0.0314     & 0.0465 & \bf 0.0177 \\
    \bottomrule
  \end{tabular}
\end{table*}

To further assess the behavioral fidelity of \mbox{Gen-C}, we conduct a \emph{random perturbation experiment} in which nodes and edges are randomly added or removed from ground-truth graphs following the proposed canonical ordering (\secref{sec:training-preprocessing}); features are randomly sampled from a uniform distribution. We use four semantic-oriented metrics capturing temporal and relational dynamics: (i) \emph{action transition KLD} and \emph{location transition KLD} (sequential consistency); (ii) \emph{joint action-location KLD} (co-occurrence distributions between actions and locations); and (iii) \emph{interaction participation rate} (proportion of agents engaged in at least one social interaction). As shown in~\tabref{tab:perturbation}, Gen-C consistently outperforms the random baseline (Perturb) across both datasets. The model achieves significantly lower KLD scores and maintains interaction participation rates closely aligned with the ground-truth (GT). These findings highlight the value of a trained model over random perturbations, showing that Gen-C effectively captures the semantic regularities underlying agent behaviors.

\begin{table}[h]
  \caption{Random perturbation comparison.}
  \label{tab:perturbation}
  \centering
  \footnotesize
  \setlength{\tabcolsep}{4.25pt}
  \begin{tabular}{l|ccc|ccc}
    \toprule
    \multirow{2}{*}[-0.1ex]{\bf Metric} & \multicolumn{3}{c|}{\textbf{University Campus}} & \multicolumn{3}{c}{\textbf{Train Station}} \\
    & Perturb & Gen-C & GT & Perturb & Gen-C & GT \\
    \midrule
    Act. Trans. $\downarrow$ & 0.584 & \textbf{0.250} & - & 0.679 & \textbf{0.239} & - \\
    Loc. Trans. $\downarrow$ & 0.566 & \textbf{0.135} & - & 0.767 & \textbf{0.427} & - \\
    Joint KLD $\downarrow$   & 0.899 & \textbf{0.248} & - & 1.121 & \textbf{0.391} & - \\
    Interact Rate & 0.176 & \textbf{0.667} & 0.699 & 0.168 & \textbf{0.627} & 0.662 \\
    \bottomrule
  \end{tabular}
\end{table}

%%%%%%%%%%%%%%%%%%%%%%%%%%%%%%%%%%%%%%%%%%%%%%%%%%%%%%%%%%%%%%%%%%%%%%%%
\subsection{Rationale for a Dedicated Learned Generator}
\label{sec:llm-only}

\begin{figure}[h]
    \centering
    \includegraphics[width=\linewidth]{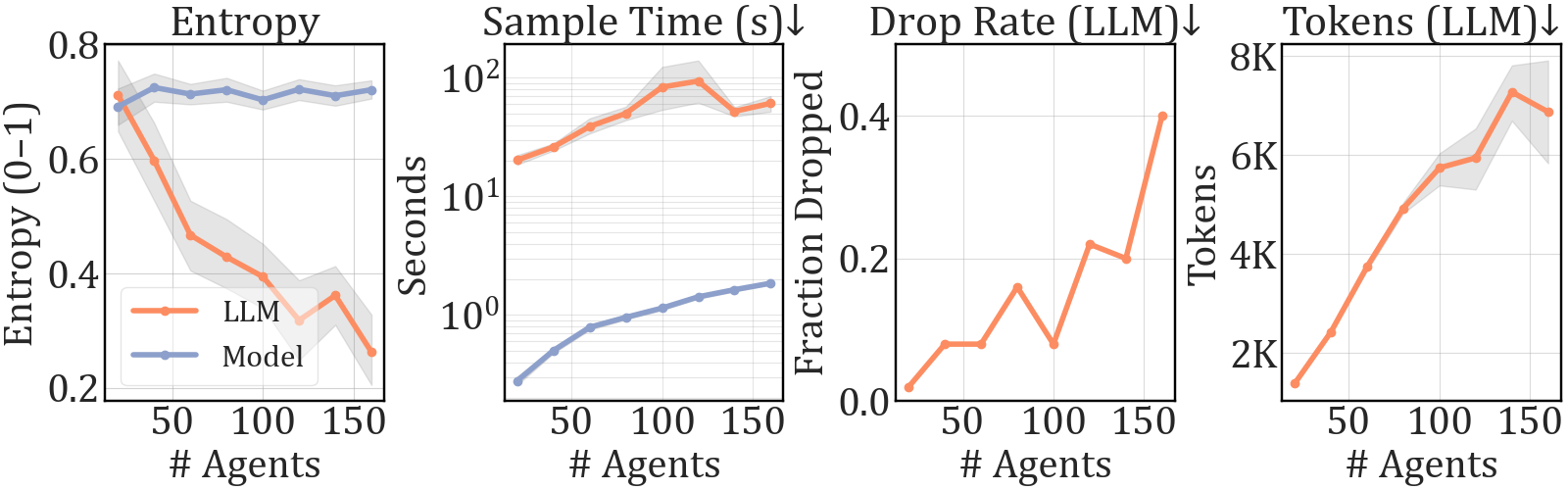}
    \caption{vanilla LLM vs Gen-C under increasing agent scale.}
    \label{fig:llm_vs_model}
\end{figure}

\noindent\textbf{Scalability, Validity, and Cost.}
We compare \emph{vanilla LLM} generation against our learned generator (\emph{Gen-C}) on University Campus under increasing agent counts $A_{count}\in[20,160]$. For consistency we proportionally scale the requested event range and the LLM’s max response tokens, and evaluate $50$ seeds per setting while reusing the same $S_{in}$ and a fixed environment designed per seed. The LLM uses the $Q2$ query (\secref{sec:quering-scenario}), updating agents/events/tokens per setting; Gen-C samples subgraphs from its prior and composes them until the target $A_{count}$ is reached. For each run we report four metrics: (i) \emph{action-sequence entropy} \(H_{\text{seq}}\) (normalized entropy over per-agent action sequences), (ii) \emph{inference time} (LLM: end-to-end completion; Gen-C: sampling\,+\,validation over composed subgraphs), (iii) \emph{token response usage (LLM only)}, and (iv) \emph{validity (LLM only)} as the \emph{drop rate}, i.e., the fraction of seeds whose realized agent count falls outside a tolerance \([0.8,1.2]\) of the target. In~\figref{fig:llm_vs_model}, we report means for metrics with $95\%$ confidence intervals.

As $A_{count}$ grows, the LLM shows a monotonic decrease in $H_{\text{seq}}$ (agents follow increasingly similar plans), a rising drop rate (more failures to meet the target $A_{count}$, and sharply increasing token usage and latency. In contrast, Gen-C maintains stable action sequence diversity and low, slowly increasing inference time. Smaller LLMs (\texttt{gpt\mbox{-}4.1\mbox{-}mini} and \texttt{gpt\mbox{-}4.1\mbox{-}nano}), show high drop rates ($0.32$ and $0.94$) even at baseline $A_{count}=20$. Overall, we show that a dedicated learned generator preserves diversity and reliability at scale, whereas prompt-based LLM generation degrades in both quality and efficiency under the same conditioning, revealing the need of a learned generator.

\begin{figure*}[t]
    \centering
    \includegraphics[width=\linewidth]{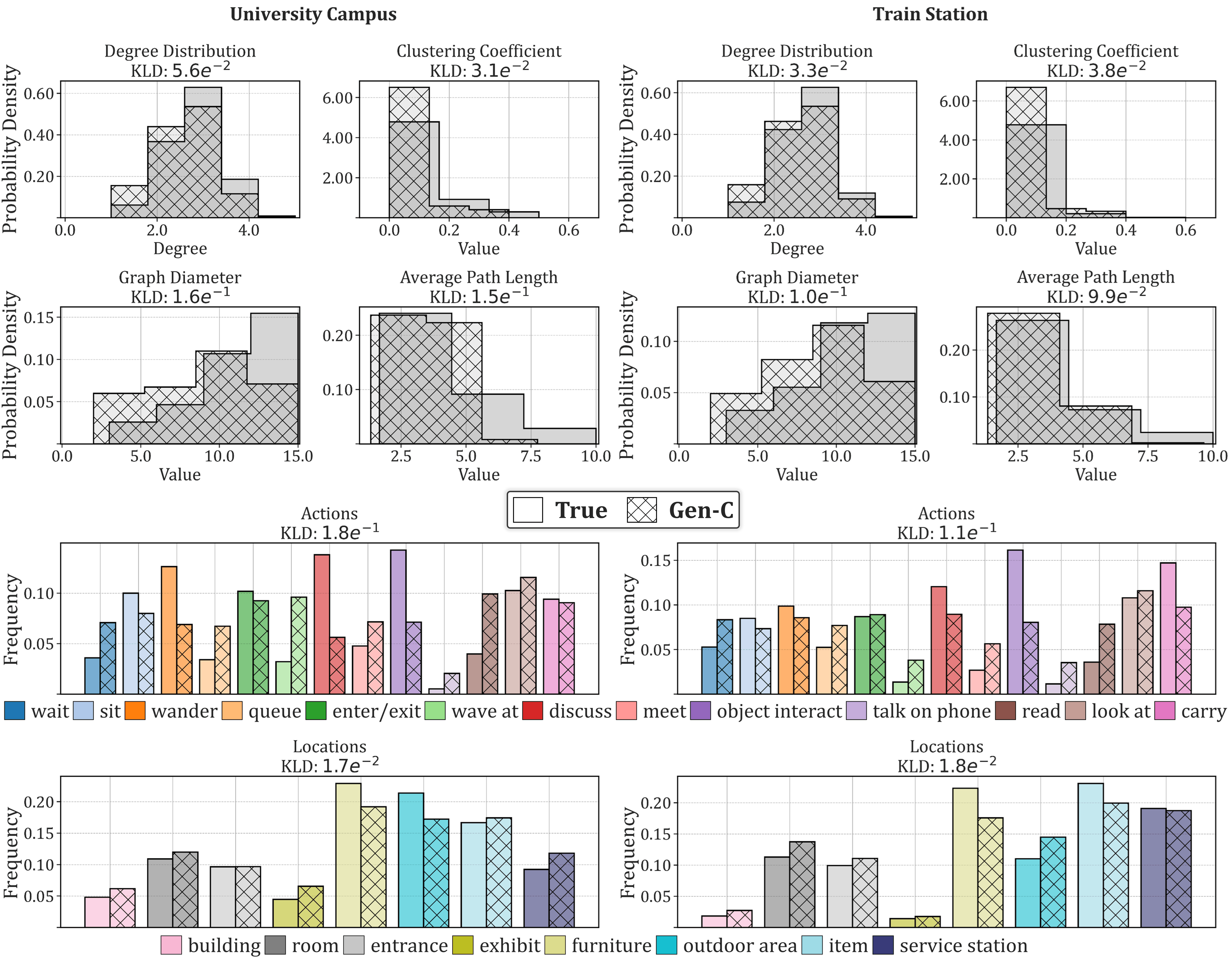}
    \caption{Graph structure and feature reconstruction quality on unseen data using samples from the conditional prior. For structure, we report various graph metrics, while for features we compare label frequencies.}
    \label{fig:statistics}
\end{figure*}

\noindent\textbf{Prompt Engineering.}
LLM-based crowd scenario generation currently relies on hand-crafted rules and constraints, creating a heavy configuration burden for non-experts. We employ a schema-guided query enforcing hard constraints (e.g., agent/event limits, diversity, no overlaps). Achieving validity requires (i) detailed instructions with few-shot examples, (ii) a custom JSON schema with post-validators, (iii) tuning of decoding parameters, and (iv) stepwise prompting with retries. These settings are brittle: small changes in scene scale and context often demand re-tuning, making the process error-prone; prompt details are in the supplementary material. In contrast, our learned generator needs only a short description and a few intuitive scalars (\#agents, \#nodes, \#events, action-frequency priors). Without prompt engineering, it sustains high validity, fidelity, and coverage with predictable latency and cost, enabling novice users to easily populate scenes with diverse crowds.

%%%%%%%%%%%%%%%%%%%%%%%%%%%%%%%%%%%%%%%%%%%%%%%%%%%%%%%%%%%%%%%%%%%%%%%%
\subsection{Quantitative Results}
\label{sec:quantitative}

\textbf{Reconstruction Quality.}  
We test the reconstruction quality of Gen-C by sampling from the conditional prior $p(Z \mid C)$ on testing set, using the input conditions of each ground-truth scenario. Our evaluation considers both (i) graph structure and (ii) node features. For structure, we report standard graph statistics, as previously introduced in~\secref{sec:ablation}. For features, we compare the distributions of action and location labels from $\mathcal{A}ct$ and $\mathcal{L}oc$.  

As shown in~\figref{fig:statistics}, we compute the KLD between ground-truth and reconstructed graph distributions; lower values indicate better alignment. VGAE-S accurately recovers structural statistics, while VGAE-F achieves strong feature reconstruction with label distributions close to the ground truth. Performance is consistent across datasets with different dynamics. For example, in the Train Station dataset ``wait'' is more common and ``meet'' less so compared to University Campus, reflecting context-specific behaviors such as travelers waiting for trains versus students meeting on campus. Similarly, ``service station'' location appears more frequently in Train Station, corresponding to information booths or arrival/departure boards, whereas outdoor areas are more prevalent in the University Campus scenario. Additionally, ``object interact'' and ``discuss'' stand out in both datasets; the former capturing context-specific actions (e.g., eating in a campus cafeteria or grabbing a ticket at a station), and the latter reflecting a common social behavior.

\begin{table*}[t]
    \caption{Top-4 action-location pairs predicted by LLM and Gen-C per theme.}
    \label{tab:top_interactions}
    \footnotesize
    \centering
    \setlength{\tabcolsep}{6pt}
    \begin{tabular}{ll}
      \toprule
      \multicolumn{2}{c}{\textbf{University Campus}} \\
      \textbf{LLM} &
      sit@furniture~$=0.08$, enter/exit@entrance~$=0.07$, wander@outdoor area~$=0.07$, discuss@furniture~$=0.05$ \\
      \textbf{Gen-C} &
      sit@furniture~$=0.06$, enter/exit@entrance~$=0.05$, wander@outdoor area~$=0.05$, wave at@outdoor area~$=0.04$ \\
      \midrule
      \multicolumn{2}{c}{\textbf{Train Station}} \\
      \textbf{LLM} &
      enter/exit@entrance~$=0.08$, sit@furniture~$=0.08$, look at@item~$=0.07$, object interact@service station~$=0.06$ \\
      \textbf{Gen-C} &
      enter/exit@entrance~$=0.07$, look at@item~$=0.06$, sit@furniture~$=0.06$, queue@service station~$=0.05$ \\
      \bottomrule
    \end{tabular}
\end{table*}

Additionally, scenario differences arise not only from action frequencies but also from their joint patterns with locations. For each theme, we analyze the joint distribution of actions and locations by aggregating all observed \texttt{action@location} pairs and computing their empirical frequencies for both LLM generations and Gen-C predictions. We select the top-k pairs, normalize their counts by the total occurrences, and report the top-4 per theme in~\tabref{tab:top_interactions}. The results show that: (i) our model preserves the dominant interaction patterns, and (ii) the prevalent pairs vary meaningfully across themes, reflecting contextual differences, while maintaining behavioral diversity.

\noindent\textbf{Latent Space Analysis.}
We assess how closely graphs sampled from the model prior match the training distribution using a \emph{fixed evaluator}: encoders trained only on training data. For each theme (University Campus \& Train Station), we embed the training set $T$ and predicted set $P$ into latent spaces $Z_T$ and $Z_P$, derived from encoder posterior means $\mu$ for structure and features. We compute the \emph{Fréchet Inception Distance (FID)}~\cite{Heusel2017GANs} and \emph{Maximum Mean Discrepancy (MMD)}~\cite{gretton2012kernel}: FID compares first- and second-order moments of the latent distributions, while MMD captures higher-order and local differences via a kernel-based similarity. Taken together, they provide complementary measures of alignment between the real and generated spaces, and are defined as follows:
\begin{equation}
\begin{gathered}
\mathrm{FID} = \lVert \mu_T - \mu_P \rVert_2^2 + \mathrm{Tr}\!\left(\Sigma_T + \Sigma_P - 2(\Sigma_T \Sigma_P)^{1/2}\right), \\
\mathrm{MMD}^2 = \mathbb{E}_{TT}[k_\gamma(x,x')] + \mathbb{E}_{PP}[k_\gamma(y,y')] - 2\,\mathbb{E}_{TP}[k_\gamma(x,y)], \\
k_\gamma(x,y) = \exp(-\gamma\|x-y\|_2^2).
\end{gathered}
\end{equation}

As shown in~\tabref{tab:latent}, FID and MMD remain low (FID: 0.53-9.29; MMD: 0.030-0.078) across both themes, with \textit{Train Station (TS)} aligning closer to the training distribution than \textit{University Campus (UC)}. The structure model (S) yields lower FID and feature model (F) lower MMD: structure latents match \emph{global moments}, while feature latents preserve \emph{local neighborhoods}. We also run cross-domain checks by \emph{mixing} model/data. Specifically, we use the model trained on \textit{University Campus} to generate samples on \textit{Train Station} and vice versa. We show that both FID (Mix) and MMD (Mix) degrade, reflecting reduced alignment across domains; this confirms that Gen-C learns appropriate latent spaces, capturing unique structure and dynamics of each domain.

\begin{table}[b]
  \caption{Latent space metrics.}
  \label{tab:latent}
  \centering
  \footnotesize
  \setlength{\tabcolsep}{6pt}
  \begin{tabular}{cc|cc|cc}
    \toprule
    \multicolumn{2}{c|}{\textbf{Setup}} & \bf FID$\downarrow$ & \bf MMD$\downarrow$ & \bf FID (Mix)$\downarrow$ & \bf MMD (Mix)$\downarrow$ \\
    \midrule
    \multirow{2}{*}{\shortstack[l]{\bf UC}} 
      & S & 1.9503 & 0.0779 & 2.9205 & 0.0986 \\
      & F & 9.2861 & 0.0319 & 17.5777 & 0.0913 \\
    \midrule
    \multirow{2}{*}{\shortstack[l]{\bf TS}}
      & S & 0.5341 & 0.0575 & 3.9793 & 0.0757 \\
      & F & 7.6735 & 0.0296 & 16.0035 & 0.1091 \\
    \bottomrule
  \end{tabular}
\end{table}

\begin{figure}[h]
  \centering
  \includegraphics[width=\linewidth]{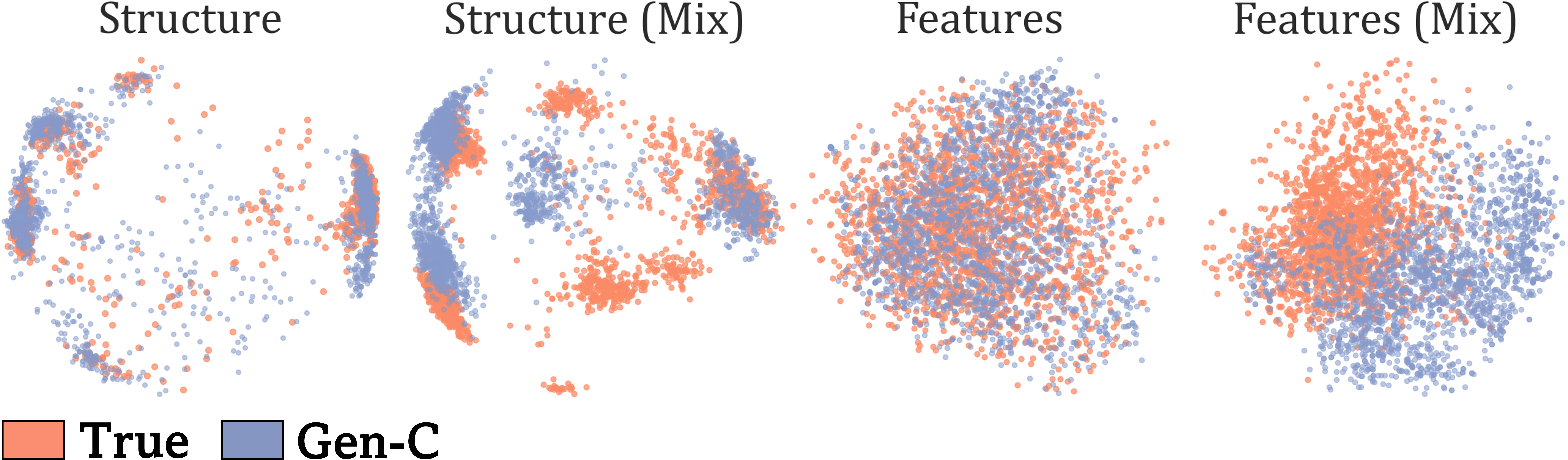}
  \caption{2K latent vectors for Train Station.}
  \label{fig:latent}
\end{figure}

Moreover, in \figref{fig:latent}, we plot 2K true and predicted latent vectors via PCA for VGAE-S (Structure) and VGAE-F (Features) on Train Station theme. In-domain, true and predicted latents largely co-locate. We also show cross-domain “mix-and-match” (Mix) projections, which exhibit reduced overlap and clear domain shifts, consistent with degraded FID/MMD in \tabref{tab:latent}. The mixed-structure space retains some overlap, as similar structures span themes, whereas the feature latent space shows a strong mismatch, clearly indicating that node features are tightly coupled to the input context. Overall, both metrics and embeddings suggest that our learned generator produces crowd scenario graphs that are well aligned with the training distribution in the evaluator space.

%%%%%%%%%%%%%%%%%%%%%%%%%%%%%%%%%%%%%%%%%%%%%%%%%%%%%%%%%%%%%%%%%%%%%%%%
\subsection{Qualitative Results}
\label{sec:qualitative}

\begin{figure*}[t]
    \centering
    \includegraphics[width=\linewidth]{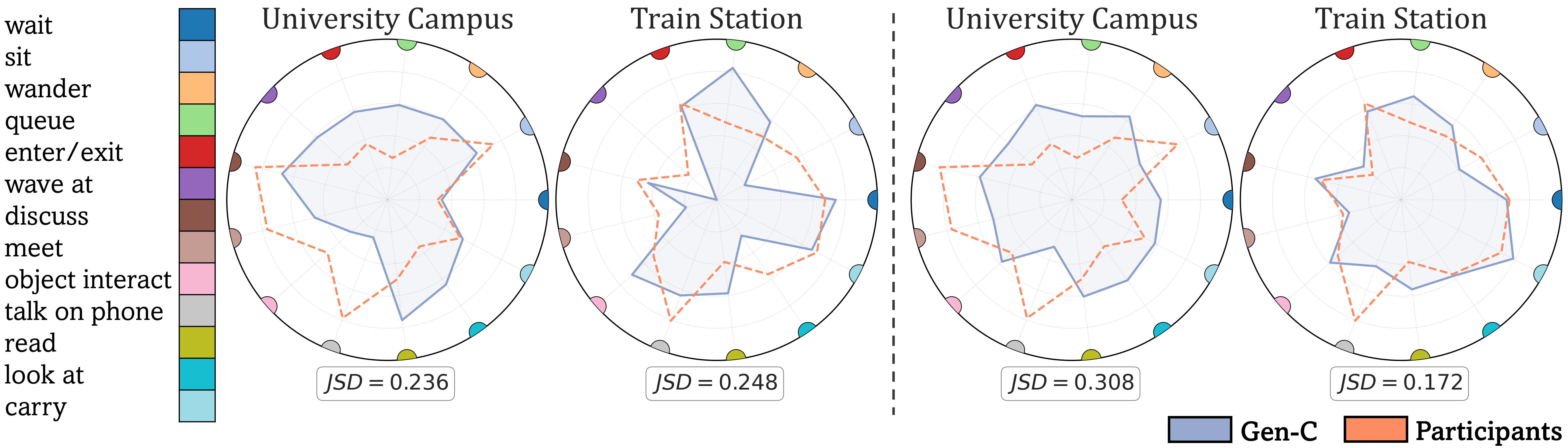}
    \caption{(Left) Comparison of action probabilities for the \emph{specific} user study scenarios. (Right) Comparison of action probabilities for the \emph{general} user study scenarios.}
    \label{fig:user_study}
\end{figure*}

\noindent\textbf{User Study.}
To assess the semantic plausibility of the generated crowd behaviors and their alignment with real-world conditions, we conducted a user study ($N=29$) comparing Gen-C predicted action distributions against human expectations. Participants were recruited through the authors’ professional networks and the Prolific platform~\cite{prolific} and completed the study via an anonymous online questionnaire. Expertise in crowd dynamics varied across respondents, with scores centered near the midpoint of the 7-point Likert scale ($M=3.50$, $SD=1.44$), and participant ages ranged from 18 to 64 years. For each of the two themes, participants reviewed seven textual descriptions (14 in total) of specific scenarios ($S_{in}$) and were asked to select the top-5 most likely actions from $\mathbf{\mathcal{A}ct}$, ranking them from most to least common; actions were presented in a random order to avoid bias. Additionally, to evaluate the model's ability to capture broader environmental ``rules'', we collected responses for four abstract, context-free queries (e.g., \textit{``You are standing on a train platform. What actions do you expect people around you to be doing?''}). Details of the user study and its questions appear in the supplementary material.

We prioritized a text-based evaluation to rigorously isolate the model’s behavioral logic from visual artifacts. In video-based assessments, particularly in multi-agent settings, user perception is often strongly influenced by rendering quality, animation smoothness, and the fidelity of motion and interactions. By removing these confounding factors, the selected assessment avoids the substantial overhead of constructing scenario-specific 3D environments and enables a focused evaluation of the semantic plausibility of the generated scenarios. Figure~\ref{fig:user_study} shows the alignment of actions between Gen-C predictions and user responses for both the (a) \emph{specific} and (b) \emph{general} scenarios. 
Specifically, for (a), we run our model on the 14 specific scenarios used in the study (20 runs per $S_{in}$) and aggregate the resulting action probabilities. The model demonstrates robust alignment with human expectations across both themes, achieving low Jensen-Shannon Divergence scores ($JSD \approx 0.24$). For (b), we compare the model's global behavior (aggregated across the full test set) against human expectations for abstract, context-free queries.
Notably, the results highlight the model's ability to capture distinct environmental ``rules'': it shows stronger convergence in the structured, rule-bound Train Station ($JSD=0.172$) compared to the unstructured, high-entropy University Campus ($JSD=0.308$), correctly reflecting the wider behavioral variance inherent to open social spaces; additional details are provided in the supplementary material.

\noindent\textbf{Comparison with Baseline Crowd Models.}
As discussed, existing crowd models primarily address local navigation and collision avoidance, and thus operate at a different layer of the simulation stack than our approach. Recent text-guided crowd synthesis methods~\cite{ji2024text} focus on group-level, goal-driven navigation, while frameworks such as GREIL-Crowds~\cite{greilcrowds} and CEDRL~\cite{Panayiotou2025CEDRL} introduce some degree of behavioral variation but their behaviors are limited to coarse categories like moving, idling, or loosely grouped motion.
The interactions between agents typically emerge implicitly from low-level navigation dynamics, while interactions with the environment are minimal or absent. Moreover, these methods are designed to generate short-horizon trajectories rather than explicit, structured behavior sequences, and cannot be conditioned on specific scenario-level descriptions beyond the general environment type (e.g., a university campus or a shopping street). In contrast, Gen-C targets the high-level decision-making and planning layer of a crowd simulator, explicitly generating scenario-conditioned sequences of semantic actions and interactions that define agent-agent and agent-environment behavior over time. Thus, Gen-C is intended to \emph{complement} existing crowd simulation approaches. As a result, our comparison with baseline models should be interpreted as a qualitative and coarse-grained analysis of behavioral diversity rather than a direct performance benchmark. We refer readers to the supplementary video for simulations produced by both the baseline models and our method.

\begin{figure*}[t]
    \centering
    \includegraphics[width=\linewidth]{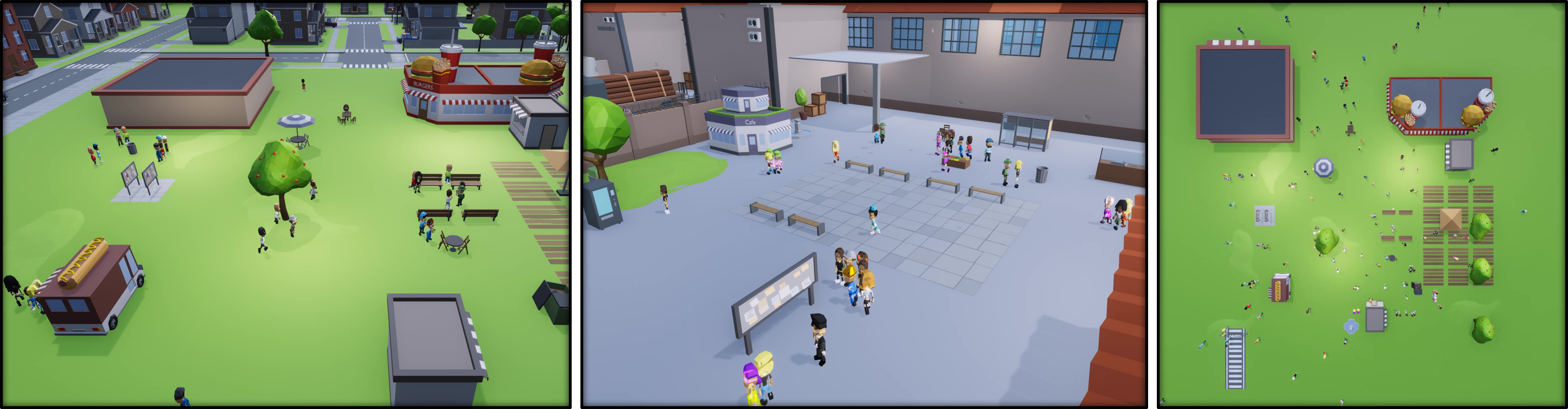}
    \caption{Our model can populate environments with rich, engaging crowd behaviors. Agents form queues for food or coffee, read announcements, wait for trains, converse over lunch, wander around, and more.}
    \label{fig:demo}
\end{figure*}

\noindent\textbf{Rendered Results.}
In~\figref{fig:demo}, we present rendered results across both themes; scenarios are simulated and visualized in Unity~\cite{unity}, with character animations manually assigned based on the high-level action each agent is performing. Gen-C generates semantically plausible behaviors, including agents queuing for food or coffee, reading announcements, waiting for trains, chatting over lunch, meeting friends, wandering through open areas, and more. These examples demonstrate the model’s ability to produce heterogeneous yet coherent multi-agent actions and interactions. Additional qualitative results are included in the supplementary video.

\section{Discussion and Future Work}
\label{sec:discussion}
We introduce \emph{Gen-C}, a framework for populating virtual scenes with multiple agents that perform diverse, high-level actions and interactions from a short textual prompt. Our framework couples a schema-guided LLM with two learned latent spaces (graph structure \& features) to synthesize crowd scenario graphs that drive agents behavior. In contrast to traditional crowd systems focused on local navigation, Gen-C targets the \emph{semantic} layer, modeling activities, locations, and social interactions, to generate plausible, varied, and scalable scenarios.

Our current implementation does not support modeling long-term agent intentions, agents cannot switch actions mid-execution, and action durations are sampled from manually defined distributions. Additionally, behaviors are constrained to a predefined list of high-level actions, constraining behavioral expressiveness and requiring retraining for new actions; environmental context is also limited to a simple positional map with constraint location types. These limitations reflect our focus on high-level planning rather than fine-grained control, and highlight several opportunities for future research. Implementing memory or belief states could enable agents to reason over past interactions and produce more coherent long-term behavior. Incorporating geometry-aware and physical feasibility constraints like density, traversability maps, and proximity penalties would ensure spatially consistent behaviors, while expanding actions via hierarchical taxonomies and hybrid datasets would boost adaptability. We would also like to integrate Gen-C with existing crowd simulators, enabling high‑level plans to guide low‑level navigation policies and thereby bridging semantic planning with physical motion.
Finally, we plan to explore combining latent spaces learned from heterogeneous datasets (e.g., indoor and outdoor scenes) using techniques such as distillation or continual learning, enabling adaptive and extensible crowd generation across diverse settings. We view our work as a step toward scalable, data-driven population of virtual worlds with rich, human-like behavior.

%%%%%%%%%%%% Supplementary Methods %%%%%%%%%%%%
%\footnotesize
%\section*{Methods}

%%%%%%%%%%%%% Acknowledgements %%%%%%%%%%%%%
%\footnotesize
%\section*{Acknowledgements}

%%%%%%%%%%%%%%   Bibliography   %%%%%%%%%%%%%%
%\normalsize
\begingroup
\setlength{\bibsep}{0pt}
\begin{scriptsize}
\bibliography{references}
\end{scriptsize}
\endgroup

%%%%%%%%%%%%%%%%   End   %%%%%%%%%%%%%%%%
%\end{multicols}  % Method B for two-column formatting (doesn't play well with line numbers), comment out if using method A
\end{document}